\documentclass[pra,aps,amssymb,twocolumn,noshowpacs]{revtex4-1}
\usepackage{color}
\usepackage{graphicx}
\usepackage{hyperref}
\usepackage{amsmath}
\usepackage{latexsym}
\usepackage{amssymb}
\usepackage{mathrsfs}
\usepackage{mathtools}
\usepackage{layout}
\usepackage{verbatim}
\usepackage{multirow}
\usepackage{bm}
\usepackage{amsfonts,epsfig}
\usepackage{rotating}

\begin{document}

\title{Transition slow-down by Rydberg interaction of neutral atoms and a fast controlled-{\footnotesize NOT} quantum gate}
\date{\today}
\author{Xiao-Feng Shi}
\affiliation{School of Physics and Optoelectronic Engineering, Xidian University, Xi'an 710071, China}

\begin{abstract}
  Exploring controllable interactions lies at the heart of quantum science. Neutral Rydberg atoms provide a versatile route toward flexible interactions between single quanta. Previous efforts mainly focused on the excitation annihilation~(EA) effect of the Rydberg blockade due to its robustness against interaction fluctuation. We study another effect of the Rydberg blockade, namely, the transition slow-down~(TSD). In TSD, a ground-Rydberg cycling in one atom slows down a Rydberg-involved state transition of a nearby atom, which is in contrast to EA that annihilates a presumed state transition. TSD can lead to an accurate controlled-{\footnotesize NOT}~({\footnotesize CNOT}) gate with a sub-$\mu$s duration about $2\pi/\Omega+\epsilon$ by two pulses, where $\epsilon$ is a negligible transient time to implement a phase change in the pulse and $\Omega$ is the Rydberg Rabi frequency. The speedy and accurate TSD-based {\footnotesize CNOT} makes neutral atoms comparable~(superior) to superconducting~(ion-trap) systems.

\end{abstract}
\maketitle

\section{introduction}
There are exciting advances in Rydberg atom quantum science recently~\cite{PhysRevLett.85.2208,Lukin2001,Saffman2010,Saffman2016,Weiss2017,Firstenberg2016,Adams2020,Browaeys2020} because of the feasibility to coherently and rapidly switch on and off the strong dipole-dipole interaction. Such interaction enables simulation of quantum many-body physics~\cite{Gurian2012,Tretyakov2017,DeLeseleuc2019,Schaub2015,Labuhn2016,Zeiher2016,Zeiher2017,Bernien2017,DeLeseleuc2018prl,Guardado-Sanchez2018,Kim2018prl,Keesling2019,Ding2020,Borish2020}, probing and manipulation of single photons~\cite{Dudin2012,Peyronel2012,Firstenberg2013,Li2013,Gorniaczyk2014,Baur2014,Tiarks2014,Li2016pan,Busche2017,Ripka2018,Liang2018,Li2019,Thompson2017}, large-scale entanglement generation~\cite{Ebert2015,Omran2019}, and quantum computation~\cite{Ebert2015,Omran2019,Wilk2010,Isenhower2010,Zhang2010,Maller2015,Zeng2017,Levine2019,Graham2019,Madjarov2020,Jau2015,Levine2018,Picken2018,Tiarks2019,Jo2019}. To date, however, most effort focused on the effect of excitation annihilation~(EA) proposed in~\cite{PhysRevLett.85.2208,Lukin2001} and reviewed in~\cite{Saffman2010,Saffman2016,Weiss2017,Firstenberg2016,Adams2020,Browaeys2020}, although there are other categories as summarized in Table~\ref{table1}. Belonging to the Rydberg blockade regime, EA involves single-atom Rydberg excitation and hence is not sensitive to the fluctuation of interaction. Besides EA, one can also excite both qubits to Rydberg states~\cite{PhysRevLett.85.2208}, or explore the antiblockade regime~\cite{Ates2007,Amthor2010}, or use the resonant dipole-dipole flip~\cite{Thompson2017,DeLeseleuc2019}. These latter processes, however, involve two-atom Rydberg excitation and are sensitive to the fluctuation of qubit separation that can substantially reduces the fidelity of a quantum control by using them~\cite{Jo2019}. It is an open question whether there is another means other than EA to explore the Rydberg blockade regime for efficient and high-fidelity quantum control.

Here, we study an unexplored transition slow-down~(TSD) effect of the dipole-dipole interaction between Rydberg atoms. When the state of one atom oscillates back and forth between ground and Rydberg states, its Rydberg interaction does not block a Rydberg-involved state swap in a nearby atom, but slows it down, and the fold of slow-down is adjustable. The resulted TSD denotes the slow-down of the state transfer of one of the two atoms although the collective Rabi frequency is enhanced due to the many-body effect. Albeit appeared as a slow-down, a controlled TSD on the contrary can drastically speed up certain crucial element in a quantum computing circuit, such as the controlled-{\footnotesize NOT}~({\footnotesize CNOT}) which is the very important two-qubit entangling gate in the circuit model of quantum computing in both theory~\cite{Nielsen2000,Williams2011,Ladd2010,Shor1997,Bremner2002,Shende2004} and experiment~\cite{Peruzzo2014,Debnath2016}. TSD is a new route toward efficient, flexible, and high-fidelity quantum control over neutral atoms because it is implemented in the strong blockade regime which is robust to the fluctuation of interactions.

\begin{table*}[ht]
  \centering
  \begin{tabular}{|c|c|c|c|c|c|c|}
    \hline
      \multicolumn{2}{|c|}{   Category}&   \multicolumn{2}{|c|}{Blockade~($V\gg\Omega$)} &   \multicolumn{3}{|c|}{Frozen interaction}\\
    \cline{1-7}
     \multicolumn{2}{|c|}{  Feature}&   \multicolumn{2}{|c|}{Robust to fluctuation of $V$} &   \multicolumn{3}{|c|}{Sensitive to fluctuation of $V$}\\
      \cline{1-7}
     \multicolumn{2}{|c|}{   Type of $V$}&   \multicolumn{2}{|c|}{\begin{tabular}{c}Dipole-dipole; Van der Waals\end{tabular}} &   \multicolumn{2}{|c|}{Van der Waals}& Dipole-dipole \\
    \cline{1-7}
    \multicolumn{2}{|c|}{  Application}&TSD &  Excitation annihilation&   Phase shift & Antiblockade& \begin{tabular}{c}State flip\end{tabular}  \\
    \cline{1-7}
     \multicolumn{2}{|c|}{ Theoretical proposal} &Here&   \cite{PhysRevLett.85.2208,Lukin2001} & \cite{PhysRevLett.85.2208}& \cite{Ates2007}&  (Intrinsic) \\
      \cline{1-7}
  \multirow{3}{*} {\begin{turn}{90}{ Realized}\end{turn}} &{\footnotesize CNOT};~Toffoli&  \multirow{2}{*} {\large{}}& \begin{tabular}{c}\cite{Isenhower2010,Zhang2010,Maller2015,Zeng2017,Levine2019,Graham2019,Tiarks2019};~\cite{Levine2019}  \end{tabular}   &   &  &\\
      \cline{2-2}\cline{4-7}
&   Entanglement &  \multirow{2}{*} {\large{}}& \begin{tabular}{c}\cite{Ebert2015,Omran2019,Wilk2010,Isenhower2010,Zhang2010,Maller2015,Jau2015,Zeng2017,Levine2018,Picken2018,Levine2019,Graham2019,Madjarov2020}, for photons:\cite{Tiarks2019}\end{tabular}   &    \cite{Jo2019}& &\\
      \cline{2-2}\cline{4-7}
    & \begin{tabular}{c}Many-body;~Optics\end{tabular}  &  & \begin{tabular}{c}\cite{Schaub2015,Labuhn2016,Zeiher2016,Zeiher2017,Bernien2017,DeLeseleuc2018prl,Guardado-Sanchez2018,Kim2018prl,Keesling2019,Ding2020,Borish2020};~\cite{Dudin2012,Peyronel2012,Firstenberg2013,Li2013,Gorniaczyk2014,Baur2014,Tiarks2014,Li2016pan,Busche2017,Ripka2018,Liang2018,Li2019}\end{tabular}   &   &\begin{tabular}{c} \cite{Ates2007,Amthor2010} \end{tabular} &\begin{tabular}{c}\cite{Gurian2012,Tretyakov2017,DeLeseleuc2019};~\cite{Thompson2017}\end{tabular} \\
   \hline    
  \end{tabular}
  \caption{ Summary of quantum science and technology based on two-body Rydberg interaction of neutral atoms~(references are not complete but representative). There are in general two categories, the strong blockade regime and the frozen interaction regime, where the former was widely studied because of its robustness to fluctuation of interactions. \label{table1}  }
\end{table*}

Though Rydberg atoms have kindled the flame for the ambition to large-scale quantum computing~\cite{PhysRevLett.85.2208,Lukin2001,Saffman2010,Saffman2016,Weiss2017}, and recent experiments demonstrated remarkable advances~\cite{Wilk2010,Isenhower2010,Zhang2010,Maller2015,Jau2015,Zeng2017,Levine2018,Picken2018,Levine2019,Graham2019,Madjarov2020,Tiarks2019,Jo2019}, further progress toward neutral-atom quantum computing is hindered by the difficulty to prepare a fast and accurate {\footnotesize CNOT}. This is partly because each of those {\footnotesize CNOT} gates was realized via combining an EA-based controlled-Z~($C_Z$) and a series of single-qubit gates~\cite{Isenhower2010,Zhang2010,Maller2015,Zeng2017,Graham2019,Levine2019}, leading to {\footnotesize CNOT} durations dominated by single-qubit operations~(e.g., over $4~\mu$s in~\cite{Levine2019,Graham2019}). In contrast, an TSD-based {\footnotesize CNOT} has a duration about $2\pi/\Omega+\epsilon$ by only two Rydberg pulses, i,e., needs no single-qubit rotations, where $\epsilon$ is a transient moment to implement a phase change in the pulses and $\Omega$ the Rydberg Rabi frequency. Using values of $\Omega$ and $\epsilon$ from~\cite{Graham2019,Levine2019}, the TSD-based {\footnotesize CNOT}~(duration $\sim0.3~\mu$s) would be orders of magnitude faster than those in~\cite{Graham2019,Levine2019}, which means that a neutral-atom {\footnotesize CNOT} can be much faster than the {\footnotesize CNOT}~(or ground Bell-state gate) by trapped ions~\cite{Cirac1995,Sorensen1999,Ballance2015,Ballance2016,Gaebler2016}~(notice that the fast ion-trap gates in~\cite{Schafer2018,Zhang2020} are phase gates). Although still inferior to superconducting circuits~\cite{You2005,Devoret2013} where {\footnotesize CNOT} gate times can be around $50$~ns~\cite{Barends2014}, the TSD-based {\footnotesize CNOT} is applicable in scalable neutral-atom platforms that are ideal for long-lived storage of quantum information in room temperatures.

\section{A two-state case}

The simplest model of TSD consists of two nearby atoms, each pumped by a laser pulse that induces a transition between a ground state $|1\rangle$ and a Rydberg state $|r\rangle$. The Rabi frequency is $\Omega_{\text{c(t)}}$ for the control~(target) qubit, and the dipole-dipole interaction $V$ in $|rr\rangle$ is assumed large compared to $\Omega_{\text{c(t)}}$ so that $|rr\rangle$ is not populated. For Rydberg interaction of the van der Waals type, $V$ is limited because in this regime the native dipole-dipole interaction should be much smaller than the energy gaps between nearby two-atom Rydberg states~\cite{Walker2008}, and thus the qubit spacing should be large enough. On the other hand, a direct dipole-dipole interaction can be huge for high-lying Rydberg states although it is no longer a pure energy shift. In this sense, the residual blockade error of the order $\sim \Omega_{\text{c(t)}}^2/V^2$~\cite{Saffman2005} can be negligible in the strong dipole-dipole interaction regime as long as the two-atom spacing is beyond the LeRoy radius.

\begin{figure}
\includegraphics[width=2.50in]
{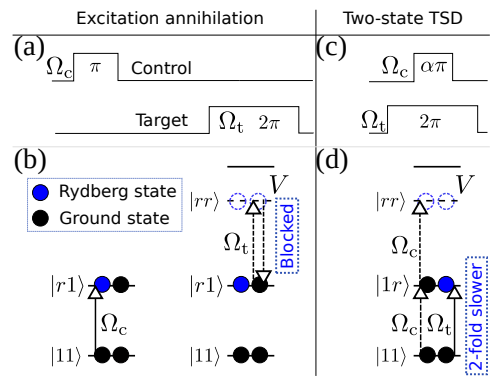}
 \caption{Comparison between excitation annihilation~(EA) and transition slow-down~(TSD) in two states. Black and blue~(gray) circles represent ground and Rydberg states, respectively. Hollow circles denote Rydberg states not populated. (a) The first and second pulses in EA are sent to the control and target qubits, respectively, both for the transition between a ground state $|1\rangle$ and a Rydberg state $|r\rangle$. (b) State evolution starting from the two-qubit state $|11\rangle$ in EA. The blockade effect is in the second pulse, where the transition $|r1\rangle\leftrightarrow|rr\rangle$ is annihilated when $\Omega_{\text{t}}\ll V$. (c) In a two-state TSD, one pulse is applied to the target qubit with duration $2t_\pi$, within which another pulse is applied to the control qubit with duration $t_\pi$, where $t_\pi=\pi/\Omega_{\text{t}}$. The delay between the pulses for the two qubits shall be smaller than $t_\pi$. (d) The parameter $\alpha\equiv\Omega_{\text{c}}/\Omega_{\text{t}}$ depends on the desired extent of TSD. With $\alpha=\sqrt{15}$, the Rydberg pumping in the target qubit, $|11\rangle\rightarrow|1r\rangle$, requires a $2\pi$ pulse, i.e., twice of that when the control qubit is not pumped.   \label{figure01} }
\end{figure}

Figure~\ref{figure01} shows a contrast between EA and a two-state TSD, where the pulse sent to the control atom is applied during $t\in [t_0,~t_0+t_1)$, and that to the target atom is during $t\in[0,~2t_1)$ with $t_0<t_1\equiv \pi/\Omega_{\text{t}}$. Starting from an initial two-atom state $|11\rangle$, the wavefunction at $t=t_0$ becomes 
\begin{eqnarray}
 |\psi(t_0)\rangle &=&\cos(\Omega_{\text{t}}t_0/2) |11\rangle -i\sin(\Omega_{\text{t}}t_0/2) |1r\rangle. \label{eq01}
\end{eqnarray}
During $t\in[t_0,~t_0+t_1)$, the system Hamiltonian is
  \begin{eqnarray}
    \hat{H} &=&[ (\Omega_{\text{c}}|r1\rangle + \Omega_{\text{t}}|1r\rangle)\langle11|/2+\text{H.c.}]+\hat{H'},\label{eq02}
\end{eqnarray}
  where $\hat{H'}$ includes excitation between $\{|1r\rangle,~|r1\rangle\}$ and $|rr\rangle$ and the dipole-dipole flip from $|rr\rangle$. Focusing on the strong interaction regime, $\hat{H'}$ can be discarded because $|rr\rangle$ is not coupled~\cite{PhysRevLett.85.2208} and hence $ \hat{H}=\bar\Omega |\mathbb{R}\rangle\langle11|/2 +\text{H.c.}$, where $\bar\Omega\equiv\sqrt{\Omega_{\text{c}}^2+\Omega_{\text{t}}^2}$ and $|\mathbb{R}\rangle\equiv (\Omega_{\text{c}}|r1\rangle + \Omega_{\text{t}}|1r\rangle)/\bar\Omega$. When $\alpha\equiv\Omega_{\text{c}}/\Omega_{\text{t}}=\sqrt{15}$, we have $\bar\Omega t_1 = 4\pi$, so that the wavefunction at $t=t_0+t_1$, given by $e^{-it_1  \hat{H}}|\psi(t_0)\rangle $, is equal to Eq.~(\ref{eq01}), and it is like that nothing happens to the target qubit upon the completion of the drive in the control qubit. Then, because the pulse for the control qubit ends at $t=t_0+t_1$, the continuous pumping on the target qubit drives the state to $|1r\rangle$ at $t=2t_1$, i.e., a $2\pi$ pulse, instead of a $\pi$ pulse, completes the transition $|11\rangle\rightarrow|1r\rangle$, which corresponds to a 2-fold slow-down. 

\begin{figure}
\includegraphics[width=3.4in]
{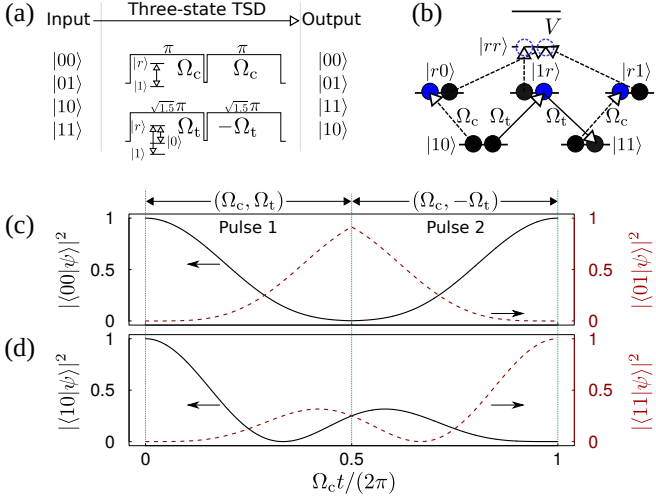}
 \caption{A three-state TSD and a fast {\footnotesize CNOT} by two pulses. (a) A three-state TSD is realized by sending to each qubit two pulses, each with equal duration $\pi/\Omega_{\text{c}}$. The Rabi frequencies are $\Omega_{\text{c}}~(\Omega_{\text{c}})$ and $\Omega_{\text{t}}~(-\Omega_{\text{t}})$ for the control and target qubits in the first~(second) pulse, where $\Omega_{\text{t}}=\sqrt{1.5}\Omega_{\text{c}}$. (b) Relevant transitions for $|10\rangle\rightarrow|11\rangle$. Transitions involving $|00\rangle$ and $|01\rangle$ are not shown because they do not involve dipole-dipole interaction. That the arrows point left or right does not mean change of angular momentum but only for clarifying the protocol.  (c) The input state $|00\rangle$ transitions back to itself because of the spin echo by $\Omega_{\text{t}}\rightarrow-\Omega_{\text{t}}$. (d) With the same pumping, the input state $|10\rangle$ transitions to $|11\rangle$ at the end of the pulse sequence. The time for implementing the phase change between the two pulses is not shown for brevity~(see text). The duration of the two pulses is $2\pi/\Omega_{\text{c}}=\sqrt{6}\pi/\Omega_{\text{t}}$, corresponding to a $\sqrt{3}$-fold slow-down because the state swap $|0\rangle\rightarrow|r\rangle\rightarrow|1\rangle$ in the target qubit can complete by a pulse duration $\sqrt{2}\pi/\Omega_{\text{t}}$ if no TSD is used.  \label{figure03} }
\end{figure}

\section{A fast {\footnotesize CNOT}}

To show the strength of TSD in quantum information, we would like to consider TSD in three states. This is because the two-state TSD ends up in a Rydberg state which is not stable, while quantum information shall be encoded with two qubit states $|0\rangle$ and $|1\rangle$ in the more stable ground manifold. For frequently used rubidium and cesium atoms, $|0\rangle$ and $|1\rangle$ can be chosen from the two hyperfine-split ground levels with a frequency difference of several gigahertz.

Consider a ground-Rydberg-ground transition chain $|0\rangle\leftrightarrow|r\rangle\leftrightarrow|1\rangle$, i.e., a state swap between the two long-lived qubit states via a metastable Rydberg state $|r\rangle$. The three-state TSD is implemented by pumping the control and target qubits with respective Rabi frequencies $\Omega_{\text{c}}$ and $\Omega_{\text{t}}$ for the same duration, with transition $|1\rangle\xleftrightarrow{\Omega_{\text{c}}}|r\rangle$ for the control qubit and $|0\rangle\xleftrightarrow{\Omega_{\text{t}}}|r\rangle\xleftrightarrow{\Omega_{\text{t}}}|1\rangle$ for the target qubit. This requires a setup capable of pumping both qubit states to a common Rydberg state in the target qubit, which is feasible as experimentally demonstrated for two-atom entanglement about a decade ago~\cite{Wilk2010} and for $W$-state preparation in atom ensembles later~\cite{Ebert2015}. Because $|0\rangle$ is not pumped in the control qubit, the Hamiltonian is $\hat{\mathcal{H}}_{c0}(\Omega_{\text{t}})=\Omega_{\text{t}}(|00\rangle+|01\rangle)\langle 0r|/2+$H.c. for the input states $|00\rangle$ and $|01\rangle$, where the subscript $c0$~($c1$) denotes that the Hamiltonian applies when the input state for the control qubit is $|0(1)\rangle$. For the remaining input states, we consider the ordered basis $\{|1r\rangle,|r1\rangle,|r0\rangle,|11\rangle,|10\rangle \}$ for the following Hamiltonian
\begin{eqnarray}
\hat{\mathcal{H}}_{c1}(\Omega_{\text{c}},\Omega_{\text{t}})&=&\frac{1}{2} \left(\begin{array}{ccccc}
0&0&0& \Omega_{\text{t}}&\Omega_{\text{t}}\\
0&0&0& \Omega_{\text{c}}&0\\
0&0&0&0& \Omega_{\text{c}}\\
\Omega_{\text{t}}& \Omega_{\text{c}}&0&0&0\\
\Omega_{\text{t}}&0& \Omega_{\text{c}}&0&0
  \end{array}\right),
  \label{eq03}
\end{eqnarray}
where  we ignore the two-atom Rydberg state $|rr\rangle$ with reasons shown below Eq.~(\ref{eq02}). By diagonalizing Eq.~(\ref{eq03}) one has
\begin{eqnarray}
 |10\rangle &=& (|\mathscr{R}_4\rangle-|\mathscr{R}_3\rangle+|\mathscr{R}_2\rangle-|\mathscr{R}_1\rangle )/2,\nonumber\\
 |11\rangle &=& (|\mathscr{R}_4\rangle-|\mathscr{R}_3\rangle-|\mathscr{R}_2\rangle+|\mathscr{R}_1\rangle )/2,\label{eq04}
\end{eqnarray}
where $|\mathscr{R}_k\rangle$ with $k=1-4$ are four eigenvectors of Eq.~(\ref{eq03}), with eigenvalues $(\Omega_{\text{c}},~-\Omega_{\text{c}},~\overline\Omega,-\overline\Omega)/2$, respectively~[the fifth eigenvector does not enter Eq.~(\ref{eq04})], where $\overline{\Omega}\equiv\sqrt{\Omega_{\text{c}}^2+2\Omega_{\text{t}}^2}$ and
\begin{eqnarray}
  |\mathscr{R}_{1,2}\rangle&=&\frac{1}{2}[|r1\rangle-|r0\rangle\pm(|11\rangle-|10\rangle)]  ,
  \label{eq05}\\
  |\mathscr{R}_{3,4}\rangle&=&\frac{1}{2\overline\Omega}[\Omega_{\text{c}}(|r1\rangle+|r0\rangle)\pm\overline\Omega(|11\rangle+|10\rangle)+2\Omega_{\text{t}}|1r\rangle].\nonumber
\end{eqnarray}

We proceed to describe an accurate and exceedingly fast {\footnotesize CNOT} by a spin-echo assisted TSD. The sequence consists of two pulses, each with duration $\pi/\Omega_{\text{c}}$ and condition $|\Omega_{\text{t}}|/\Omega_{\text{c}}=\sqrt{6}/2$, shown in Fig.~\ref{figure03}(a). A $\pi$ phase change~(requiring a time $\epsilon$) is inserted between the pulses sent to the target qubit so as to induce spin echo, where $\epsilon$ can be around $10$~ns~\cite{Levine2019}. The spin echo suppresses the state swap of the target qubit if the control qubit is initialized in $|0\rangle$~\cite{Shi2018prapp2,Shi2020}; but when the control qubit is initialized in $|1\rangle$, it is pumped to $|r\rangle$ which results in TSD in the target qubit, i.e., the transition $|0\rangle\rightleftharpoons|r\rangle\rightleftharpoons|1\rangle$ in the target qubit, which can occur with a pulse duration $\sqrt{2}\pi/\Omega_{\text{t}}$ if no TSD is used~\cite{Shi2018prapp}, will be slowed down by a fold of $\sqrt{3}$. The mechanism is understood in two steps. {\it First}, during $t\in[0,~t_\pi)$ with $t_\pi=\pi/\Omega_{\text{c}}$, the input state $|\psi_{c0}(0)\rangle=|00\rangle$ or $|01\rangle$ evolves according to $e^{-it\mathcal{H}_{c0}(\Omega_{\text{t}})}|\psi_{c0}\rangle$, and after a $\pi$ phase change to the Rydberg Rabi frequency which may require a finite transient time $\epsilon$, the state evolution becomes $e^{-i(t-t_\pi-\epsilon)\mathcal{H}_{c0}(-\Omega_{\text{t}})}|\psi_{c0}(t_\pi)\rangle$ during the second pulse, leading to $e^{-it\mathcal{H}_{c0}(-\Omega_{\text{t}})t_\pi}|\psi_{c0}(t_\pi)\rangle=e^{-it\mathcal{H}_{c0}(-\Omega_{\text{t}})t_\pi}e^{-it\mathcal{H}_{c0}(\Omega_{\text{t}})t_\pi}|\psi_{c0}(0)\rangle=|\psi_{c0}(0)\rangle$ at the end of the sequence. {\it Second}, according to Eq.~(\ref{eq05}), the first pulse during $t\in[0,~t_\pi)$ evolves the input state $|\psi_{c1}(0)\rangle=|10\rangle$ according to
\begin{eqnarray}
\frac{1}{2} (e^{\frac{it\overline\Omega}{2}}|\mathscr{R}_4\rangle-e^{\frac{-it\overline\Omega}{2}}|\mathscr{R}_3\rangle+e^{\frac{it\Omega_{\text{c}}}{2}}|\mathscr{R}_2\rangle -e^{-\frac{it\Omega_{\text{c}}}{2}}|\mathscr{R}_1\rangle),\label{eq06}
\end{eqnarray}
which becomes
\begin{eqnarray}
|\psi_{c1}(t_\pi)\rangle &=&\frac{1}{2} (i|\mathscr{R}_1\rangle+i|\mathscr{R}_2\rangle+|\mathscr{R}_3\rangle-|\mathscr{R}_4\rangle )\label{eq07}
\end{eqnarray}
at $t=t_\pi$ because $(t_\pi\Omega_{\text{c}},~t_\pi\overline\Omega)=(\pi,~2\pi)$. In Eq.~(\ref{eq07}), the eigenvectors $|\mathscr{R}_j\rangle$, $j=1-4$, are defined in Eq.~(\ref{eq05}) with Rabi frequencies $(\Omega_{\text{c}},~\Omega_{\text{t}})$ during $t\in[0,~t_\pi)$. During $t\in t_\pi+\epsilon+[0,~t_\pi)$, the Hamiltonian has Rabi frequencies $(\Omega_{\text{c}},~-\Omega_{\text{t}})$ with the eigenvectors $|\mathscr{R}_{1,2}\rangle$ the same as in Eq.~(\ref{eq05}), but the expressions for $|\mathscr{R}_{3,4}\rangle$ become   
\begin{eqnarray}
    |\mathscr{R}_{3,4}'\rangle&=&\frac{1}{2\overline\Omega}[\Omega_{\text{c}}(|r1\rangle+|r0\rangle)\pm\overline\Omega(|11\rangle+|10\rangle)-2\Omega_{\text{t}}|1r\rangle].\nonumber\\
  \label{eq08}
\end{eqnarray}
Using Eqs.~(\ref{eq05}) and~(\ref{eq08}) we cast Eq.~(\ref{eq07}) into
\begin{eqnarray}
|\psi_{c1}(t_\pi)\rangle =e^{-it_\pi \mathcal{H}_{c1}(\Omega_{\text{c}},-\Omega_{\text{t}})}|\psi_{c1}(0)\rangle,
  \label{eq-key}
\end{eqnarray}
i.e., the state at $t=t_\pi$ is like if Rabi frequencies $(\Omega_{\text{c}},~-\Omega_{\text{t}})$ were used during the first pulse. Equation~(\ref{eq-key}) is the key feature enabling a fast {\footnotesize CNOT}. After the second pulse, the state at $t=2t_\pi+\epsilon$ becomes $e^{-it_\pi\mathcal{H}_{c1}(\Omega_{\text{c}},-\Omega_{\text{t}})}|\psi_{c1}(t_\pi)\rangle$ that can be rewritten as $e^{-2it_\pi\mathcal{H}_{c1}(\Omega_{\text{c}},-\Omega_{\text{t}})}|\psi_{c1}(0)\rangle$ thanks to Eq.~(\ref{eq-key}), which further reduces to $|11\rangle$ according to Eqs.~(\ref{eq04}) and~(\ref{eq06}). A similar analysis shows that if the initial state is $|11\rangle$, it maps to $|10\rangle$ upon the completion of the pulse. So the following map is realized
\begin{eqnarray}
\{|00\rangle, |01\rangle, |10\rangle, |11\rangle\}\rightarrow\{|00\rangle, |01\rangle, |11\rangle, |10\rangle\}\label{eq09}
\end{eqnarray}
which is the standard {\footnotesize CNOT}. A numerical simulation of the population evolution by $\mathcal{H}_{c0}$ and $\mathcal{H}_{c1}$ in Eq.~(\ref{eq03}) is shown in Figs.~\ref{figure03}(c) and~\ref{figure03}(d) for the input states $|00\rangle$ and $|10\rangle$, respectively, where the transient time $\epsilon$ is ignored for brevity.

High fidelity is possible with the TSD-based {\footnotesize CNOT}. To estimate the intrinsic fidelity limited by Rydberg-state decay and Doppler dephasing, we consider atomic levels and Rydberg laser Rabi frequencies in recent experiments~\cite{Graham2019,Levine2019}, and numerically found that the fidelity of the TSD-based {\footnotesize CNOT} or Bell state would reach $99.9\%$~($99.8\%$) with experimentally affordable effective temperatures $T_a=5~(15)~\mu$K of qubit motion~\cite{Picken2018,Zeng2017,Graham2019}. As detailed in Appendix~\ref{app02}, this estimate assumes that the field for $|0\rangle\rightarrow|r\rangle$ and that for $|1\rangle\rightarrow|r\rangle$ in the target qubit copropagate, leading to opposite phases in the two transitions of the chain $|0\rangle\xrightarrow{\Omega_{\text{t}}e^{itkv_t}}|r\rangle\xrightarrow{\Omega_{\text{t}}e^{-itkv_t}}|1\rangle$, where $k$ is the wavevector and $v_t$ the speed of the target qubit along the propagation direction of light. Then, the state transfer from $|0\rangle$ to $|1\rangle$~(or the reverse) picks up two opposite phases $\pm tkv_t$ which partly suppresses the dephasing. Fortunately, our method is not strictly dependent on the above dephasing-resilient configuration. For example, for the worst case of counterpropagating fields so that the sequential state transfer from $|0\rangle$ to $|1\rangle$ picks up a total phase $2tkv_t$, simulation in Appendix~\ref{appendixA} shows that fidelity over $99.6\%$ is achievable for $T_a\sim15~\mu$K in room temperatures. The robustness against the Doppler dephasing benefits from the avoidance of shelving the control atom on the Rydberg level in free flight during pumping the target atom as required in an EA-based {\footnotesize CNOT}~\cite{Muller2009,Shi2018prapp}.

The above analysis assumes $V/2\pi=500$~MHz corresponding to an interatomic distance $l=3~\mu$m. When $V/2\pi=100$~MHz with larger $l=4.6~\mu$m (where crosstalk can be negligible~\cite{Graham2019}), the fidelity would be $99.8\%$~($99.7\%$) with $T_a=5~(15)~\mu$K in room temperatures~(see Appendix~\ref{SMsec03} for details). This benefits from that our {\footnotesize CNOT} is not sensitive to the change of interaction $V$ in the strong blockade regime. Finally, with fluctuation~(of relative Gaussian width $\sigma$) of the two Rabi frequencies for the two transitions $|0(1)\rangle\leftrightarrow|r\rangle$ in the target qubit, very small extra error $\in(0.02,~0.4)\%$ arises for $\sigma\in(1,~5)\%$ as detailed in Appendix~\ref{appendixC}. So, high fidelity is achievable since the power fluctuation of Rydberg lasers can be well suppressed~\cite{Levine2018,Levine2019}. 

\section{Discussion and conclusions}

The {\footnotesize CNOT} in Eq.~(\ref{eq09}) is implemented within a short time $2\pi/\Omega_{\text{c}}=\sqrt{6}\pi/\Omega_{\text{t}}$ plus a transient moment $\epsilon$ for a phase change between the two pulses, where $\epsilon$ can be negligible as in~\cite{Levine2019}. Although a phase twist is used in both~\cite{Levine2019} and here, our TSD method is drastically different in physics  from the method of~\cite{Levine2019} as discussed in Appendix~\ref{appendixD}. Note that Eq.~(\ref{eq09}) does not depend on any single-qubit gates if one starts from Eq.~(\ref{eq03}). If $s$ or $d$-orbital Rydberg states are used, two-photon transitions can be applied which leads to ac Stark shifts for the ground and Rydberg states. As shown in Appendix~\ref{appendixE}, these shifts are annulled in Eq.~(\ref{eq03}) which is achievable by choosing appropriate ratio between the detuning at the intermediate state and the magnitudes of the fields for the lower and upper transitions~\cite{Maller2015}. 

Our {\footnotesize CNOT} gate duration can be $\lesssim0.3~\mu$s with Rydberg Rabi frequencies~($2\pi\times(3.5\sim4.6)$~MHz) like those realized in~\cite{Graham2019,Levine2019} that showed {\footnotesize CNOT} gate durations orders of magnitude longer than here. Previous {\footnotesize CNOT} gates~\cite{Zhang2010,Maller2015,Zeng2017,Graham2019,Levine2019,Isenhower2010} required two or more single-qubit rotations to convert $C_Z$ to {\footnotesize CNOT}, which was usually achieved with two-frequency Raman light~\cite{Isenhower2010,Zhang2010,Zeng2017,Levine2019}, or with microwave driving assisted by Stark shift of laser fields~\cite{Graham2019,Maller2015}. High-fidelity realizations of such gates came with long gate durations~\cite{Xia2015,Wang2016,Graham2019,Levine2019}, although they can also be carried out rapidly in proof-of-principle experiments~\cite{Yavuz2006,Jones2007} where high fidelity is not concerned. The atom traps have finite lifetimes~\cite{Saffman2016} and a faster protocol can lead to more {\footnotesize CNOT} cycles before the qubit arrays should be reloaded. Because a practical quantum processor tackling real problems uses a series of quantum gates including {\footnotesize CNOT}~\cite{Nielsen2000,Williams2011,Ladd2010,Shor1997,Bremner2002,Shende2004,Peruzzo2014,Debnath2016}, certain computation tasks can be executed only with fast enough {\footnotesize CNOT} in each cycle of laser cooling and loading of atom array. The studied speedup of the minimal functional quantum circuit shows a possibility to confront the problems of finite lifetime of atom traps which limits the capability of a neutral-atom quantum processor. Moreover, concerning the ratio between the coherence time and the {\footnotesize CNOT} gate time, the TSD method makes neutral-atom systems competitive with superconducting and ion-trap systems as compared in detail in Appendix~\ref{appendixF}.

In conclusion, we have explored another effect of Rydberg blockade by dipole-dipole interaction in neutral atoms, namely, the effect of transition slow-down~(TSD). We show that TSD can speed up a neutral-atom quantum computer by proposing an exceedingly fast TSD-based {\footnotesize CNOT} realized by two pulses separated by a short transient moment for changing the phase of the pulse. The exotic TSD is an excellent character of dipole-dipole interactions other than the well-known excitation annihilation effect that can push Rydberg-atom quantum science to another level.

\section*{ACKNOWLEDGMENTS}
The author thanks Yan Lu for useful discussions. This work is supported by the National Natural Science Foundation of China under Grant No. 11805146 and Natural Science Basic Research plan in Shaanxi Province of China under Grant No. 2020JM-189.

\appendix{}
\section{Gate fidelity}\label{appendixA}
 Before analyzing the gate error, we would like to emphasize that TSD is particularly useful for speeding up the minimal functional quantum circuit, namely, the {\footnotesize CNOT}. The {\footnotesize CNOT} protocols by EA in previous methods involve multiple switchings of the external control fields which brings extra complexity to the experimental implementation, elongate the gate duration, and introduces extra errors from the fluctuation of the control fields and from the intrinsic atomic Doppler broadening. Previous effort for suppressing these errors includes a $C_Z$ gate based on quantum interference~\cite{Shi2019,Levine2019}, but going from $C_Z$ to the more useful {\footnotesize CNOT}~(or to create Bell states in the ground-state manifold) still needs several single-qubit operations~\cite{Graham2019,Levine2019,Isenhower2010,Zhang2010,Maller2015,Zeng2017}. For example, the first experimental neutral-atom {\footnotesize CNOT} needed five or seven pulses~\cite{Isenhower2010}, and the recent {\footnotesize CNOT} in~\cite{Levine2019} used two~(four) pulses in the control~(target) qubit besides several short pulses for phase compensation. The central procedure of the {\footnotesize CNOT} in~\cite{Levine2019} is via combining (i) a two-qubit $C_Z$-like gate by two pulses of duration about $2.7\pi/\Omega$ with a phase change inserted between the pulses that requires an extra transient time $\epsilon$, (ii) a short pulse to compensate an intrinsic phase to recover a $C_Z$ from the $C_Z$-like gate, and (iii) two single-qubit rotations of duration $\pi/\Omega_{\text{hf}}$ ($\approx2~\mu$s therein) in the target qubit, where $\Omega$ is the Rydberg laser Rabi frequency and $\Omega_{\text{hf}}$ is the hyperfine laser Rabi frequency between the two qubit states. Page 4 of Ref.~\cite{Levine2019} indicates that the $C_Z$-like gate needs $0.4~\mu$s, and it had $(\Omega_{\text{hf}},~\Omega)/(2\pi)=(0.25,~3.5)~$MHz~(see page 1 and 2 of~\cite{Levine2019}), thus $\epsilon=0.4~\mu$s$-2.732\pi/\Omega\approx9.7$~ns; we assume such a fast phase change time here. Figure 3(d) of Ref.~\cite{Levine2019} presents a {\footnotesize CNOT} sequence with durations equal to those of two $X(\pi/2)$, one $X(\pi)$, and the $C_Z$-like, i.e., $2\times \pi/(2\Omega_{\text{hf}}) + \pi/\Omega_{\text{hf}}+2.7\pi/\Omega$, and thus the {\footnotesize CNOT} sequence needs about $4.4~\mu$s therein~(the actual gate durations should be larger when accounting for gaps between pulses therein). In contrast, the TSD-based {\footnotesize CNOT} needs a duration about $\sqrt{6}\pi/\Omega_{\text{t}}$ which is only $0.27~(0.35)~\mu$s with $\Omega_{\text{t}}/(2\pi)=4.6~(3.5)$~MHz from Ref.~\cite{Graham2019}~(Ref.~\cite{Levine2019}). 

 One may image that if the single-qubit rotations in Refs.~\cite{Graham2019,Levine2019} are implemented by the transition chain $|0\rangle\leftrightarrow|r\rangle\leftrightarrow|1\rangle$, then their gate durations can also be small. But there will be problems in this assumption. This is because the single-qubit rotations necessary to transform $C_Z$ to {\footnotesize CNOT} are two $X(\pi/2)$ gates that transfer $\{|0\rangle,~|1\rangle\}$ to $\{|0\rangle-i|1\rangle,~-i|0\rangle+|1\rangle\}/\sqrt{2}$~(see Fig.~5(a) of~\cite{Graham2019} or Fig.~3(d) of~\cite{Levine2019}). But the pumping by $\hat{\mathbb{H}}(\Omega_{\text{t}})=\Omega_{\text{t}}(|0\rangle+|1\rangle)\langle r|/2+$H.c can not achieve this since one can easily prove that starting from $|\varphi\rangle=|0\rangle$, the populations in $\{|0\rangle,~|r\rangle,~|1\rangle\}$ will be $\frac{1}{2}\{2\cos^4\frac{\theta}{2} ,\sin^2\theta,~2\sin^4\frac{\theta}{2}\}$, where $\theta=\frac{t\Omega_{\text{t}}}{\sqrt{2}}$. This means that there is no way to use the resonant transition chain $|0\rangle\leftrightarrow|r\rangle\leftrightarrow|1\rangle$ for the $X(\pi/2)$ rotation. On the other hand, one may also imagine a succession of a $\pi$ pulse on $|1\rangle\rightarrow |r\rangle$, a $\pi/2$ pulse on $|r\rangle\rightarrow |0\rangle$, and a $\pi$ pulse on $|r\rangle\rightarrow |1\rangle$ can, e.g., realize a $X(\pi/2)$ rotation. However, the extra time to shelve an atom in Rydberg state leads to extra Doppler dephasing, and the frequent turning on and off of Rydberg lasers can lead to extra atom loss~\cite{Maller2015}.

 Below, we analyze gate imperfections due to the prevailing intrinsic Rydberg-state decay and Doppler broadening. These are the dominant intrinsic errors in gate operations~\cite{Graham2019} while technical issues such as laser noise is in principle not fundamental. From here to Appendix~\ref{appendixC}, we take, as an example, $^{87}$Rb qubits and consider the intermediate level $6P_{3/2}$ for Rydberg pumping, where the two detunings at $6P_{3/2}$ should be different for the two transitions $|0\rangle\xleftrightarrow{\Omega_{\text{t}}}|r\rangle$ and $|1\rangle\xleftrightarrow{\Omega_{\text{t}}}|r\rangle$. For an $s$-orbital rubidium Rydberg state with principal quantum number around $70$, the lifetime of $|r\rangle$ is about $\tau=400~(150)~\mu$s at a temperature of $4~(300)$~K by the estimate in~\cite{Beterov2009}. When the lower and upper fields counterpropagate along, e.g., $\mathbf{z}$, the wavevector is $k=2\pi(1/420.3-1/1012.7)$nm$^{-1}$~\cite{Shi2020prapp}. We assume that the two qubits are initially located at the centers of the traps at $(0,0,x_0)$ and $(0,0,0)$ respectively. The traps are usually turned off during Rydberg pumping, and the free flight of the qubits leads to time-dependent Rabi frequencies $\Omega_{\text{c}}e^{it(kv_c+x_0)}$ for the control qubit, and $(\Omega_{\text{t}}e^{itkv_t},\Omega_{\text{t}}e^{itkv_t})$ for the two transitions $|0(1)\rangle\leftrightarrow|r\rangle$ in the target qubit, where $(v_c,~v_t)$ are the projection of velocity along $\mathbf{z}$ for the control and target qubits, respectively. With a finite atomic temperature $T_{\text{a}}$, there is a finite distribution $\mathscr{D}(v_c)\mathscr{D}(v_t)$ for the speeds $(v_c,v_t)$, where $\mathscr{D}(v)$ is a Gaussian~\cite{Graham2019,Shi2020prapp}.

\subsection{Rydberg-state decay}\label{app01}
Although the gate duration, when neglecting the transient time~($<10$~ns as shown above) for phase change in the pulses, is $t_{\text{g}}=\frac{2\pi}{\Omega_{\text{c}}}$, the main decay error arises when the atoms are in Rydberg state~\cite{Saffman2005} supposing the intermediate state is largely detuned. By using the estimate in~\cite{Zhang2012}, the decay error of the TSD-based {\footnotesize CNOT} gate can be approximated as
\begin{eqnarray}
  E_{\text{decay}} &=& \frac{1}{4\tau}\int dt \Big[\sum_{|\psi(0)\rangle=|00\rangle,|01\rangle} |\langle 0r|\psi(t)\rangle|^2
    \nonumber\\
    &&+\sum_{|\psi(0)\rangle=|10\rangle,|11\rangle}\big( |\langle r0|\psi(t)\rangle|^2+|\langle r1|\psi(t)\rangle|^2\nonumber\\
    &&+|\langle 1r|\psi(t)\rangle|^2 \big)\Big],
\end{eqnarray}
which is $0.39t_{\text{g}}/\tau$ by numerical simulation. Consider a set of experimentally feasible~\cite{Graham2019} values of Rydberg Rabi frequencies $\Omega_{\text{c}}=\Omega_{\text{t}}/\sqrt{1.5}=2\pi\times3.6$~MHz, we have $E_{\text{decay}}=2.7~(7.2)\times10^{-4}$ for qubits in an environment temperature of $4~(300)$~K. Alternatively, a more detailed numerical simulation by using the optical Bloch equation in the Lindblad form with correct branching ratios~\cite{Shi2019} can predict a slightly lower $E_{\text{decay}}$ because some population can decay to qubit states that will again contribute to the gate operation.  

\subsection{Doppler dephasing}\label{app02}
For the TSD-based {\footnotesize CNOT} to be resilient to Doppler dephasing, the two fields for the lower transitions $|0\rangle\rightarrow|p\rangle$ and $|1\rangle\rightarrow|p\rangle$ shall copropagate along $\mathbf{z}$, and those for their upper transitions $|p\rangle\rightarrow|r\rangle$ shall copropagate along $-\mathbf{z}$, so that the wavevectors for $|0(1)\rangle\rightarrow|r\rangle$ have (approximately) the same value $k$. Here $|p\rangle$ is symbolic for the intermediate $6P_{3/2}$ state and one shall bare in mind that the detunings at $|p\rangle$ for the two transition chains must have a large difference; alternatively one can use different fine states in the $6p$ manifold for the two transition chains~(the numerical results in Tables~\ref{tableSM1} and~\ref{tableSM2} stay similar). Then, the transition $|0\rangle\xrightarrow{\Omega_{\text{t}}}|r\rangle\xrightarrow{\Omega_{\text{t}}}|1\rangle$ becomes 
\begin{eqnarray}
 |0\rangle\xrightarrow{\Omega_{\text{t}}e^{itkv_t}}|r\rangle\xrightarrow{\Omega_{\text{t}}e^{-itkv_t}}|1\rangle \label{conditionFie}
\end{eqnarray}
when accounting for the atom drift. The above transition mainly transfers the population between the two hyperfine states, which means that if negligible population stays at $|r\rangle$, the two phases $tkv_t$ and $-tkv_t$ add up for any moment, leading to negligible phase noise because $tkv_t-tkv_t=0$ for the population transfer from $|0\rangle$ to $|1\rangle$. However, this is not ideal since there is always some population at $|r\rangle$ during the process. Nonetheless, that there is partial phase cancellation can suppress the Doppler dephasing compared to usual cases.

For the control qubit, the transition $|1\rangle\xrightarrow{\Omega_{\text{c}}e^{itkv_c}}|r\rangle$ still has the usual Doppler dephasing. However, the pumping $|1\rangle\xrightarrow{\Omega_{\text{c}}e^{itkv_c}}|r\rangle$ in the control is immersed in the TSD and does not put much population in the Rydberg state. Numerical simulation shows $\int dt [ |\langle r0|\psi(t)\rangle|^2+|\langle r1|\psi(t)\rangle|^2]=0.31t_{\text{g}}$ for either $|\psi(0)\rangle=|00\rangle$ or $|01\rangle$ in each gate sequence.

To show the robustness of the TSD-based {\footnotesize CNOT} against the Doppler dephasing, we numerically simulate the state evolution by using $\hat{\mathcal{H}}_{c0}(\Omega_{\text{t}}e^{itk v_t})=\Omega_{\text{t}}(e^{-itkv_t}|00\rangle+e^{-itkv_t}|01\rangle)\langle 0r|/2+$H.c. for the input states $|00\rangle$ and $|01\rangle$, and
\begin{widetext}
\begin{eqnarray}
  \hat{\mathcal{H}}_{c1}(\Omega_{\text{c}}e^{itk v_c},\Omega_{\text{t}}e^{itk v_t}) &=& \frac{1}{2} \left(\begin{array}{cccccc}
  2V &\Omega_{\text{c}}e^{itk v_c}& \Omega_{\text{t}}e^{itk v_t}&\Omega_{\text{t}}e^{itk v_t}&0&0\\
\Omega_{\text{c}}e^{-itk v_c}&0&0&0& \Omega_{\text{t}}e^{itk v_t}&\Omega_{\text{t}}e^{itk v_t}\\
\Omega_{\text{t}}e^{-itk v_t}&0&0&0& \Omega_{\text{c}}e^{itk v_c}&0\\
\Omega_{\text{t}}e^{-itk v_t}&0&0&0&0& \Omega_{\text{c}}e^{itk v_c}\\
0&\Omega_{\text{t}}e^{-itk v_t}& \Omega_{\text{c}}e^{-itk v_c}&0&0&0\\
0&\Omega_{\text{t}}e^{-itk v_t}&0& \Omega_{\text{c}}e^{-itk v_c}&0&0
  \end{array}\right),
  \label{SM03}
\end{eqnarray}
\end{widetext}
for the input states $|10\rangle$ and $|11\rangle$, where Eq.~(\ref{SM03}) is written with the basis $\{|rr\rangle, |1r\rangle,|r1\rangle,|r0\rangle,|11\rangle,|10\rangle  \}$, where $V$ represents the interaction of the state $|rr\rangle$. Because of the Doppler dephasing, the gate map in the basis $\{|00\rangle, |01\rangle, |10\rangle, |11\rangle\}$ changes from the ideal form
\begin{eqnarray}
U &=& \left(
  \begin{array}{cccc}
    1& 0 & 0&0\\
    0 & 1 &0&0\\
    0 &0 & 0&1\\
    0& 0 & 1&0\\   
    \end{array} 
  \right) ,\label{cnot}
  \end{eqnarray}
to
\begin{eqnarray}
 \mathscr{U} &=& \left(
  \begin{array}{cccc}
    a& b & 0&0\\
    c & d &0&0\\
    0 &0 & e&f\\
    0& 0 & g&h\\   
    \end{array} 
  \right) ,\label{realgate01}
  \end{eqnarray}
where $a,b,c$, and $d$ can be calculated by sequentially using $\hat{\mathcal{H}}_{c0}(\Omega_{\text{t}})$ and $\hat{\mathcal{H}}_{c0}(-\Omega_{\text{t}})$ for the input states $|00\rangle$ and $|01\rangle$, while $e,f,g$, and $h$ can be calculated using, sequentially, $\hat{\mathcal{H}}_{c1}(\Omega_{\text{c}},\Omega_{\text{t}})$ and $\hat{\mathcal{H}}_{c1}(\Omega_{\text{c}},-\Omega_{\text{t}})$, for the input states $|10\rangle$ and $|11\rangle$. To study the robustness to Doppler dephasing, we would like to see errors mainly from the Doppler dephasing if the blockade error is negligible. So we adopt a large blockade interaction $V/(2\pi)=500$~MHz as in Ref.~\cite{Saffman2020}. We choose $\Omega_{\text{c}}=\Omega_{\text{t}}/\sqrt{1.5}=2\pi\times3.6$~MHz, define the rotation error by~\cite{Pedersen2007}
\begin{eqnarray}
 E_{\text{ro}} &=& 1-\frac{1}{20}\left[  |\text{Tr}(U^\dag \mathscr{U})|^2 + \text{Tr}(U^\dag \mathscr{U}\mathscr{U}^\dag U ) \right], \label{fidelityError01}
\end{eqnarray}
and evaluate the ensemble average with
\begin{eqnarray}
\overline{E_{\text{ro}}}\approx \frac{\sum_{v_c}\sum_{v_t} E_{\text{ro}}(v_c,~v_t)\mathscr{D}(v_c) \mathscr{D}(v_t) }{ \sum_{v_c}\sum_{v_t} \mathscr{D}(v_c) \mathscr{D}(v_t)},\label{roterror02}
\end{eqnarray}
where the sum is over $10^4$ sets of speeds $(v_{c},~v_t)$, where $v_{c(t)}$ applies $101$ values equally distributed from $-0.5$ to $0.5$~m/s because the atomic speed has little chance to be over $0.5$~m/s for the temperatures $T_a\leq50~\mu$K considered in this work. The approximation by Eq.~(\ref{roterror02}) has little difference from a rigorous integration. More details can be found in~\cite{Shi2020prapp} for the method of numerical simulation.

{\it Case 1.}--By using $\hat{\mathcal{H}}_{c0}(\Omega_{\text{t}}e^{itk v_t})$~(for $\{|00\rangle,|01\rangle\}$) and $\hat{\mathcal{H}}_{c1}(\Omega_{\text{c}}e^{itk v_c},\Omega_{\text{t}}e^{itk v_t})$~(for $\{|10\rangle,|11\rangle\}$) for the first pulse, and $\hat{\mathcal{H}}_{c0}(\Omega_{\text{t}}e^{itk v_t})$ and $\hat{\mathcal{H}}_{c1}(\Omega_{\text{c}}e^{itk v_c},-\Omega_{\text{t}}e^{itk v_t})$ for the second pulse, the ensemble-averaged rotation errors are given in the second row of Table~\ref{tableSM1}, with effective atomic temperatures above $T_a=5~\mu$K which was achievable in experiments~\cite{Picken2018}. From these results, one can see that the {\footnotesize CNOT} fidelity $1-E_{\text{decay}}-\overline{E_{\text{ro}}}$ can reach $99.9\%$ with qubits cooled to around $T_a=10~(5)~\mu$K in a $4~(300)~$K environment.

{\it Case 2.}--To further suppress the Doppler dephasing, we consider switching propagation directions of fields for the control qubit between the two pulses. In other words, the Hamiltonians~(for $\{|00\rangle,|01\rangle\}$) are $\hat{\mathcal{H}}_{c0}(\Omega_{\text{t}}e^{itk v_t})$ and $\hat{\mathcal{H}}_{c0}(-\Omega_{\text{t}}e^{-itk v_t})$ for the first and second pulses, respectively. The ensemble-averaged rotation errors are shown in the third row of Table~\ref{tableSM1}. From the results in Table~\ref{tableSM1}, one can see that the {\footnotesize CNOT} fidelity $1-E_{\text{decay}}-\overline{E_{\text{ro}}}$ has a slight improvement if this latter case of configuration is employed. The mechanism for the extra suppression of Doppler dephasing in this latter case lies in that the population going, e.g., from $|0\rangle$ to $|1\rangle$, obtains some phase error due to the action $\Omega_{\text{t}}e^{-it'kv_t}|1\rangle\langle r|$ and $\Omega_{\text{t}}e^{itkv_t}|r\rangle\langle 0|$; subsequently, the second pulse pumps $|1\rangle$ to $|0\rangle$ and induces some phase error due to the action $-\Omega_{\text{t}}e^{-it'kv_t}|r\rangle\langle 1|$ and $-\Omega_{\text{t}}e^{itkv_t}|0\rangle\langle r|$, which means that the phase terms between the two pulses are continuous which is better for the desired transition to occur. For case 1, the pumping from $|0\rangle$ to $|r\rangle$ has a phase term $tkv_t$ at the end of pulse 1, which becomes $-tkv_t$ when the population begins to go back at the start of pulse 2, i.e., there is a phase jump. More details about the influence on gate fidelity from atom-drift-induced phase change in Rabi frequencies can be found in~\cite{Shi2020prapp}. Since this latter case is a little involved and the increase of fidelity is marginal, it is only of interest in the future when technical issues are resolved so that a fidelity level beyond $99\%$ is technically achievable.   

 \begin{table}
  \centering
  \begin{tabular}{|c |c|c|c|c|c|c|}
    \hline
 & $T~(\mu$K) &  5&10&15 & 20 & 50 \\\hline
Case 1 &$10^4\times\overline{E_{\text{ro}}}$& $4.31$ &  $8.09$&  $11.9$ &$15.6$& $38.2$  \\\hline 
Case 2&$10^4\times\overline{E_{\text{ro}}}$ & $3.11$ &  $5.69$&  $8.26$ & $10.8$& $26.2$ \\\hline 
  \end{tabular}
  \caption{ Rotation error~(scaled up by $10^4$; excluding Rydberg-state decay) of the TSD-based {\footnotesize CNOT} by $\Omega_{c}/2\pi=3.5$~MHz with two cases, where case 2 corresponds to that the propagation directions for the fields on the target qubit are switched between the first and second pulses, while no such switching occurs in case 1. Here Eq.~(\ref{fidelityError01}) was used that accounts for the phase errors and the truth table errors.  \label{tableSM1}  }
  \end{table}

 \begin{table}
  \centering
  \begin{tabular}{|c |c|c|c|c|c|c|}
    \hline
 & $T~(\mu$K) &  5&10&15 & 20 & 50 \\\hline
Case 1 &$10^4\times\overline{E_{\text{Bell}}}$& $2.86$ &  $5.27$&  $7.67$ &$10.1$& $24.4$  \\\hline 
Case 2&$10^4\times\overline{E_{\text{Bell}}}$ & $2.57$ &  $4.67$&  $6.78$ & $8.88$& $21.4$ \\\hline 
  \end{tabular}
  \caption{ Fidelity error~(scaled up by $10^4$; excluding Rydberg-state decay) of the Bell state by TSD-based {\footnotesize CNOT}, with the same pulse sequence as used for Table~\ref{tableSM1}. $\overline{(\cdots)}$ denotes ensemble average as Eq.~(\ref{roterror02}).  \label{tableSM2}  }
  \end{table}
 
 The result in Table~\ref{tableSM1} accounts for both the population errors and the phase errors. In experiments, usually the phase errors are not important especially in the characterization of the Bell state~\cite{Levine2019,Graham2019}. Thus we continue to study the strength of TSD-based {\footnotesize CNOT} for achieving high-fidelity Bell states.
 
\subsection{High-fidelity Bell states}
A TSD-based {\footnotesize CNOT} can map the initial product state $|\psi(0)\rangle= (|00\rangle+|10\rangle)/\sqrt{2}$ to the Bell state $|\Phi\rangle= (|00\rangle+|11\rangle)/\sqrt{2}$ without resorting to extra single-qubit gates~(except the state initialization). The Rydberg superposition time is equal to that in Sec.~\ref{app01}. We follow Ref.~\cite{Levine2019} by evaluating the infidelity as $E_{\text{Bell}}=1-\langle \Phi|\rho(t)|\Phi\rangle $, where $\rho(t)=|\psi\rangle\langle\psi|$ with $|\psi\rangle$ evolved by using the pulse sequence as in Sec.~\ref{app02} for the input state $|\psi(0)\rangle$. The numerical results are given in Table~\ref{tableSM2}, which shows that in room temperatures, the Bell state can be created with a fidelity $99.88\%$ if qubits are cooled to the level of $T_a=10~\mu$K, which is affordable according to the atom cooling achieved in previous experiments~\cite{Picken2018,Zeng2017}.  

Finally, we would like to emphasize that it is not so crucial to have the Doppler-resilient configuration~(with copropagating fields) for the two transition chains in the target qubit, as discussed around Eq.~(\ref{conditionFie}). Consider the worst case with largest dephasing, i.e., if the two sets of fields counterpropagate so that Eq.~(\ref{conditionFie}) becomes $|0\rangle\xrightarrow{\Omega_{\text{t}}e^{itkv_t}}|r\rangle\xrightarrow{\Omega_{\text{t}}e^{itkv_t}}|1\rangle$, then the motional dephasing error should be larger. For example, we numerically found that the Bell-state errors in the second row (case 1) of Table~\ref{tableSM2} become $(9.72\times10^{-4},~1.90\times10^{-3},~2.82\times10^{-3},~3.74\times10^{-3},~9.22\times10^{-3})$ for $T_a=(5,~10,~15,~20,~50)~\mu$K; similar results can be found for the TSD-based CNOT. With $E_{\text{decay}}=7.2\times10^{-4}$ for qubits in an environment of $300$~K, the Bell-state fidelity would be $99.65\%$ at $T_a=15~\mu$K. This still large fidelity means that for experimental convenience, high fidelity is possible with any configuration for the propagation directions for the two sets of fields if errors other than the intrinsic Doppler dephasing and Rydberg-state decay can be avoided.

\begin{figure}
\includegraphics[width=3.4in]
{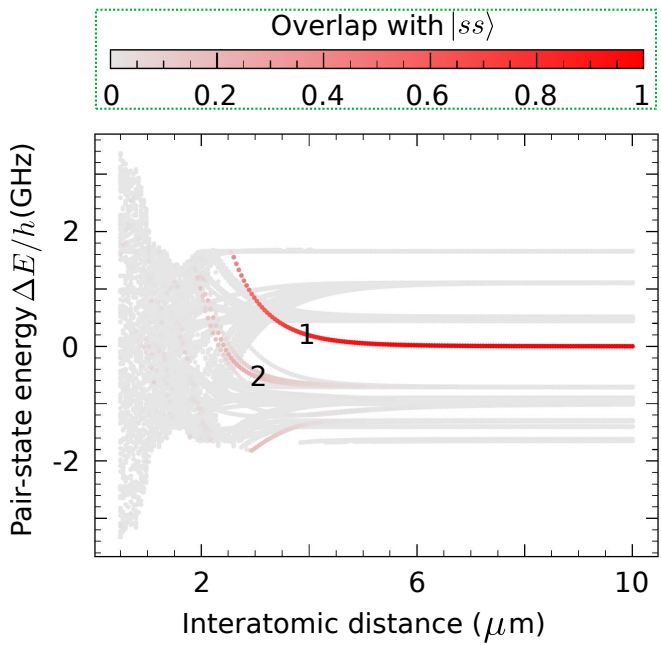}
 \caption{Energy spectrum for two $^{87}$Rb atoms lying along the quantization axis when initialized in the two-atom pair state $|ss\rangle\equiv|70S_{1/2},m_J=1/2;70S_{1/2},m_J=1/2\rangle$~\cite{Sibalic2017}. Here 1 and 2 label the two states that have largest overlap with $|ss\rangle$ and nearest to the unperturbed energy (labeled as zero).  \label{figureSM01} }
\end{figure}

\section{Interatomic distances}\label{SMsec03}
The dipole-dipole interaction $V$ is a function of interatomic distance $\mathcal{L}$. For $V$ to be large enough so that the blockade error is negligible, $\mathcal{L}$ should be small enough. In order to avoid wavefunction overlap, the distance between the nuclei of the two Rydberg atoms shall be larger than the Le Roy distance which can be calculated by using the open source library of Ref.~\cite{Sibalic2017}. For the parameters chosen as an example in this work, the Le Roy distance is $1.50~\mu$m for two rubidium atoms in the state $70S_{1/2}$, and we can consider longer interatomic distance in the range of $\mathcal{L}>2.6\mu$m. The dipole-dipole interaction will couple the two-atom state $|70S_{1/2},m_J=1/2;70S_{1/2},m_J=1/2\rangle$ to many other states. To find the interaction, we consider states $|nL;n'L'\rangle$ coupled from the initial states ($n$ and $L$ denote the principal and angular momentum quantum numbers, respectively), with $|n-70|, |n'-70|\leq 5$, and $|L|, |L'|\leq 4$ (because $ss$ states couple to $pp$ states, which couple to $d$-orbital and to $f$-orbital states, and so on), and consider energy gaps between pair states within $2\pi\times25$~GHz. After diagonalization, the energy map is shown in Fig.~\ref{figureSM01}. In Fig.~\ref{figureSM01}, the color denotes the population of the state $|70S_{1/2},m_J=1/2;70S_{1/2},m_J=1/2\rangle$ in the diagonalized state; the color can also be understood as how possible if one atom is already in the state $|70S_{1/2},m_J=1/2\rangle$, what will the chance be to populate the state if the other atom is Rydberg pumped via, e.g., the $5P_{3/2}$ intermediate state.

For the TSD to hold, we focus on the strong blockade regime where the two-atom Rydberg state shown in Fig.~\ref{figureSM01} is barely populated. In this case, the {\it smallest} energy of the diagonalized eigenstate matters, or more accurately, the state with the smallest eigenenergy that can be coupled (marked by red) plays the role. In Fig.~\ref{figureSM01}, one can find that for $\mathcal{L}\in[2.6,~4.6]\mu$m, there are mainly three eigenstates, one with a positive energy~(labeled as state 1), and the other two with negative energy. For those two that have negative energy, the one we focus on is the state with more overlap with $|ss\rangle\equiv|70S_{1/2},m_J=1/2;70S_{1/2},m_J=1/2\rangle$, whose energy is sometimes lower and sometimes higher than the energy of the other with less population in $|ss\rangle$. The two states we focus on are labeled by 1 and 2 in Fig.~\ref{figureSM01}. State 1 has the largest component in $|ss\rangle$: at $\mathcal{L}=\{2.6,~3.0,~3.6,~4.0,~4.6\}\mu$m, the amplitude overlap between state 1 and $|ss\rangle$ is $\{0.64,~0.74,~0.84,~0.9,~0.95\}$. For state 2, its overlap with $|ss\rangle$ is $\{0.49,~0.43,~0.32\}$ when $\mathcal{L}=\{2.6,~3.0,~3.6\}\mu$m and is negligible when $\mathcal{L}$ is beyond $4.0~\mu$m. The eigenenergy of these two states are listed in Table~\ref{tableSM3}, from which one can see that around $\mathcal{L}=3.0~\mu$m, the interaction $V=2\pi\times510$ can be represented by the eigenenergy of state 2, which is largest for the cases shown in Table~\ref{tableSM3}.

 \begin{table}
  \centering
  \begin{tabular}{|c|c||c|c|c|c|c|}
    \hline
&$\mathcal{L}~(\mu$m) & 2.6 & 3.0&3.6 & 4.0&4.6 \\\hline 
State 1&$V/(2\pi)$~(MHz) & 1600& 780&340& 180& 94 \\\hline
State 2&$V/(2\pi)$~(MHz) &-280&-510 &-590 & -650& -670 \\\hline
  \end{tabular}
  \caption{ Eigenenergy of the two eigenstates of the dipole-coupled two-atom states that have the largest overlap with the state $|70S_{1/2},m_J=1/2;70S_{1/2},m_J=1/2\rangle$ and that are nearest to zero, labeled as state 1 and state 2 in Fig.~\ref{figureSM01}. \label{tableSM3}  }
  \end{table}
 \begin{table}
  \centering
  \begin{tabular}{|c ||c |c|c|c|c|c|c|}
    \hline
$V/(2\pi)$~(MHz) & $T~(\mu$K) &  5&10&15 & 20 & 50 \\\hline 
50 &$10^4\times\overline{E_{\text{ro}}}$& $56.5$ &  $60.3$&  $64.1$ &$67.8$& $90.3$  \\\hline
100&$10^4\times\overline{E_{\text{ro}}}$ & $17.0$ &  $20.8$&  $24.5$ & $28.3$& $58.6$ \\\hline 
200&$10^4\times\overline{E_{\text{ro}}}$ & $7.09$ &  $10.9$&  $14.6$ & $18.4$& $41.0$ \\\hline 
300&$10^4\times\overline{E_{\text{ro}}}$ & $5.25$ &  $9.03$&  $12.8$ & $16.6$& $39.1$ \\\hline 
400&$10^4\times\overline{E_{\text{ro}}}$ & $4.61$ &  $8.39$&  $12.2$ & $15.9$& $38.5$ \\\hline 
  \end{tabular}
  \caption{ Rotation error~(scaled up by $10^4$; excluding Rydberg-state decay) of the TSD-based {\footnotesize CNOT} by $\Omega_{c}/2\pi=3.5$~MHz with different values of $V$.  \label{tableSM4}  }
  \end{table}

To have $V=2\pi\times500$~MHz, the above study shows that placing the two qubits with a distance around $\mathcal{L}=3.0~\mu$m seems necessary. With such a distance, the analysis in Ref.~\cite{Graham2019} shows that the crosstalk error is about $0.5\%$ if the waist~($1/e^2$ intensity radii) of the laser beams is $w=3.0~\mu$m. To reduce the crosstalk, it is possible to use super-Gaussian Rydberg beams, as detailed in Ref.~\cite{Gillen-Christandl2016}. Another possibility is to use higher Rydberg states so that the dipole-dipole interaction can be as large as $V=2\pi\times500$~MHz at a larger interatomic distance, and hence the laser-beam crosstalk can be avoided. 

If only commonly used Gaussian Rydberg beams are employed for a Rydberg state of principal quantum number around $n=70$, then smaller values of $V$ can be used for larger interatomic distances so as to avoid crosstalk. For the values $V/(2\pi)\in[50,~400]$~MHz, the fidelity of our gate is shown in Table~\ref{tableSM4}, which shows that with $V=2\pi\times100$~MHz, the rotation error is about $1.7\times10^{-3}$ with $T_a=5~\mu$K, and the interatomic distance should be around $\mathcal{L}=4.6\mu$m where crosstalk can be ignored safely~($\Omega'/\Omega\sim e^{-2\mathcal{L}^2/w^2}=0.009$) if the waists for the lower and upper lasers are both $w=3.0~\mu$m. Taking into account the decay error, this means that our gate would have a fidelity $99.76\%$ with $T_a=5~\mu$K in room temperatures.  

\begin{figure}
\includegraphics[width=3.4in]
{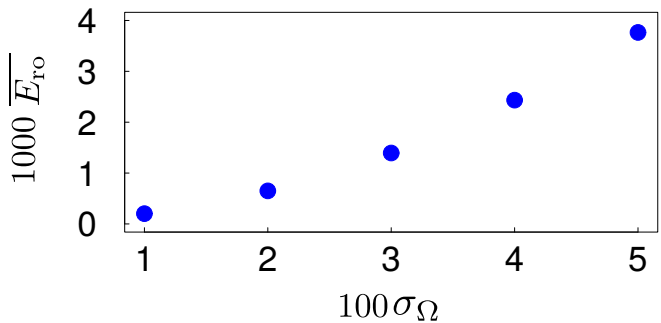}
 \caption{Fidelity error from the fluctuation in the Rabi frequencies for the two transitions $|0\rangle\leftrightarrow|r\rangle$ and $|1\rangle\leftrightarrow|r\rangle$; the fluctuation obeys a Gaussian distribution of width $\sigma_\Omega$. The error is $0.00376$ for $\sigma_\Omega=5\%$.  \label{figureSM02} }
\end{figure}

\section{Amplitude fluctuation of laser fields }\label{appendixC}
Our TSD-based {\footnotesize CNOT} requires the Rabi frequencies for the two transitions $|0\rangle\leftrightarrow|r\rangle\leftrightarrow|1\rangle$ to be equal in the target qubit. Here we investigate the impact on the gate fidelity if this condition is not satisfied.

We assume that the Rabi frequency $\Omega$ obeys a Gaussian distribution
\begin{eqnarray}
 \mathscr{G}(\Omega) &=&\frac{1}{\sigma\sqrt{2\pi}}e^{-\frac{(\Omega-\Omega_{k})^2}{2\sigma^2}}
\end{eqnarray}
around the desired value $\Omega_{k}$, where $k\in\{$t1,~t2$\}$ denotes the two channels $|0\rangle\leftrightarrow|r\rangle$ and $|1\rangle\leftrightarrow|r\rangle$. By using the Gaussian distributed Rabi frequencies in the target qubit, one can evaluate the averaged fidelity by 
\begin{eqnarray}
\overline{E_{\text{ro}}}\approx \frac{\sum_{\Omega_{\text{t1}}}\sum_{\Omega_{\text{t2}}} E_{\text{ro}}(\Omega_{\text{t1}}, \Omega_{\text{t2}})\mathscr{G}(\Omega_{\text{t1}}) \mathscr{G}(\Omega_{\text{t2}})   }{\sum_{\Omega_{\text{t1}}}\sum_{\Omega_{\text{t2}}} \mathscr{G}(\Omega_{\text{t1}}) \mathscr{G}(\Omega_{\text{t2}})   },\label{roterror07}
\end{eqnarray}
where $E_{\text{ro}}(\Omega_{\text{t1}}, \Omega_{\text{t2}})$ is evaluated by using the gate fidelity with $\Omega_{\text{t1}}$ and $\Omega_{\text{t2}}$ for the two transitions $|0\rangle\leftrightarrow|r\rangle$ and $|1\rangle\leftrightarrow|r\rangle$. In Eq.~(\ref{roterror07}), the integration is approximated by the sum over 121 sets of $(\Omega_{\text{t1}}, \Omega_{\text{t2}})$, each of which applies values $(\Omega-\Omega_{k})/\Omega_{k}\in\{\pm 5,~\pm4,~\pm 3,~\pm 2,~\pm 1,~0\}\sigma$. With $\sigma\in\{0.01,~0.02,~0.03,~0.04,~0.05\}$, the result is shown in Fig.~\ref{figureSM02}. One can see that with quite large relative fluctuation of Gaussian width $5\%$, the fidelity is still larger than $99.6\%$. There is little relation between the Doppler dephasing and fluctuation of Rabi frequency, and one can expect that the gate fidelity can be evaluated by combining Table~\ref{tableSM1} and Fig.~\ref{figureSM02} if the fluctuation of the Rabi frequencies is not quenched.

In Ref.~\cite{Levine2018}, the power fluctuation of the Rydberg lasers was suppressed below $1\%$ for preparing ground-Rydberg entanglement, which means that the Rydberg Rabi frequencies were suppressed to have relative noise below $10\%$ in Ref.~\cite{Levine2018}. More than one year later, in Ref.~\cite{Levine2019}, preparation of high-fidelity ground-state entanglement was reported by the same group, which was a significant progress. We suppose that the laser noise was even more suppressed in Ref.~\cite{Levine2019} compared to the earlier experiment in Ref.~\cite{Levine2018}, and thus we consider the relative Gaussian width of the fluctuation up to $5\%$.

\section{Comparison with other fast Rydberg gates}\label{appendixD}
It is useful to compare our{ \footnotesize CNOT} with other fast entangling methods with Rydberg atoms. The comparison focuses on physical mechanism and their application in quantum computing.  

A popular method to generate entanglement by Rydberg interactions is to use the EA mechanism~\cite{PhysRevLett.85.2208}. The standard way to use it is a three-pulse sequence for a $C_Z$ gate in the form of
\begin{eqnarray}
C_Z &=& \left(
  \begin{array}{cccc}
    1& 0 & 0&0\\
    0 & -1 &0&0\\
    0 &0 & -1&0\\
    0& 0 & 0&-1\\   
    \end{array} 
  \right) \label{czmap}
  \end{eqnarray}
in the basis $\{|00\rangle, |01\rangle, |10\rangle, |11\rangle\}$, where the blockade takes effect in the state $|11\rangle$. To use it for quantum computing in the circuit model, the $C_Z$ gate in Eq.~(\ref{czmap}) needs single-qubit gates to become a {\footnotesize CNOT}. By this method, a most recent experiment in Ref.~\cite{Graham2019} realized a {\footnotesize CNOT} of duration more than 100~$\mu$s with Rydberg Rabi frequencies~$2\pi\times4.6$~MHz because of the slow single-qubit gates. Physically, this method depends on a missed $\pi$ phase accumulation in an annihilated Rabi cycle for the input state $|11\rangle$ by the Rydberg blockade.  

The other recent experiment in Ref.~\cite{Levine2019} used detuned Rabi cycles for entanglement combined with detunings found by optimal control. The method is essentially identical to the interference method proposed in Ref.~\cite{Shi2019}, as can be easily verified by looking at the similar structure of the gate matrix~[with the same basis as in Eq.~(\ref{czmap})]
\begin{eqnarray}
C_{\text{phase}} &=& \left(
  \begin{array}{cccc}
    1& 0 & 0&0\\
    0 & e^{i\alpha} &0&0\\
    0 &0 & e^{i\alpha}&0\\
    0& 0 & 0&e^{i\beta}\\   
    \end{array} 
  \right) \label{czlikemap}
\end{eqnarray}
in Refs.~\cite{Levine2019,Shi2019}~(Ref.~\cite{Shi2019} proposed two gates among which the first is quoted here). Compared to the initial interference method, Ref.~\cite{Levine2019} used optimal-control-found parameters and phase twist in the Rydberg pumping within the blockade regime so as to realize $2\alpha-\beta=\pi$~(and thus the gate in Ref.~\cite{Levine2019} can be called a $C_Z$-like gate). Because of the detuned Rabi cycles used, there is no way to realize a gate with $\{\alpha,~\beta\}/\pi=\{N_1,N_2\}$, where $N_j$ is an integer with $j=1,2$. To use it in quantum computing, several single-qubit gates must be used to transform the gate in Ref.~\cite{Levine2019} to the {\footnotesize CNOT} so their final {\footnotesize CNOT} duration was more than $4~\mu$s with Rydberg Rabi frequencies~$2\pi\times3.5$~MHz. Without using the detuned Rabi cycles, the method in Ref.~\cite{Levine2019} can not work. The physical mechanism of Ref.~\cite{Levine2019} is that dynamical phases in detuned Rabi cycles are accumulated. 

In order to compare with the gates in Eqs.~(\ref{czmap}) and (\ref{czlikemap}) which have a diagonal form, the TSD-based {\footnotesize CNOT} in the basis of $\{|00\rangle, |01\rangle, |10\rangle, |11\rangle\}$ can be rewritten as 
\begin{eqnarray}
C_{\text{TSD}} &=& \left(
  \begin{array}{cccc}
    1& 0 & 0&0\\
    0 & 1 &0&0\\
    0 &0 & -1&0\\
    0& 0 & 0&1\\   
    \end{array} 
  \right) ,\label{cnotothermap}
\end{eqnarray}
with the basis $\{|0\overline{0}\rangle, |0\overline{1}\rangle, |1\overline{0}\rangle, |1\overline{1}\rangle\}$, where $|\overline{1}(\overline{0})\rangle=(|0\rangle\pm|1\rangle)/\sqrt{2}$. The {\footnotesize CNOT} duration would be less than $0.4~\mu$s with Rydberg Rabi frequencies~$\leq2\pi\times3.5$~MHz by the TSD method. The blockade interaction is in $|1\overline{1}\rangle$ and the pumping on the target qubit induces a transition from $|1\overline{1}\rangle$ to $|\mathscr{S}_+\rangle\equiv(\Omega_{\text{c}}|r\overline{1}\rangle+\sqrt{2}\Omega_{\text{t}}|1r\rangle)/\overline{\Omega}$ with a Rabi frequency $\overline{\Omega}=\sqrt{\Omega_{\text{c}}^2+2\Omega_{\text{t}}^2}$ while further excitation to $|rr\rangle$ is blocked. Each of the two pulses in the TSD-based {\footnotesize CNOT} sequence induces a complete $2\pi$ rotation $|1\overline{1}\rangle\rightarrow -i |\mathscr{S}_+\rangle\rightarrow -|1\overline{1}\rangle$ which leads to a $\pi$ phase change to $|1\overline{1}\rangle$. The change of the sign of $\Omega_{\text{t}}$ between the two pulses does not change this picture, and thus a total phase $2\pi$ is accumulated in $|1\overline{1}\rangle$. Since $e^{2i\pi}=1$, $|1\overline{1}\rangle$ acquires no phase term in practice. On the other hand, the input state $|1\overline{0}\rangle$ only experiences two $\pi$ pulses~(i.e., a $2\pi$ pulse) in the control qubit which results in a $\pi$ phase change to it. The pumping of the target qubit experiences spin echo for the input state $|0\overline{1}\rangle$, thus no phase appears for it. So, the physical mechanism of the gate in Eq.~(\ref{cnotothermap}) is fundamentally different from those in Eqs.~(\ref{czmap}) and (\ref{czlikemap}); in fact, the drastically different forms of the three gate maps reveal this. To sum up in one word, Eq.~(\ref{czmap}) relies on a missed $\pi$ phase change in an annihilated Rabi cycle, Eq.~(\ref{czlikemap}) relies on three phase changes in three detuned Rabi cycles, and~(\ref{cnotothermap}) relies on a $\pi$ phase change in a resonant Rabi cycle. For quantum computing, the gate in Eq.~(\ref{cnotothermap}) is exactly a {\footnotesize CNOT} in the basis of $\{|00\rangle, |01\rangle, |10\rangle, |11\rangle\}$, and thus is more useful compared to Eqs.~(\ref{czmap}) and (\ref{czlikemap}). 

Because a phase twist is used in realizing both Eq.~(\ref{czlikemap}) and Eq.~(\ref{cnotothermap}), one may guess that they are similar. But the following facts show their distinct physics: (i) detuned Rydberg pumping is used in Eq.~(\ref{czlikemap}), but resonant Rydberg pumping is used in Eq.~(\ref{cnotothermap}); (ii) three input states acquire phase terms in Eq.~(\ref{czlikemap}), but only one input state acquires a phase term in Eq.~(\ref{cnotothermap}); (iii) the population in Rydberg state can reach 1 for the fourth input state $|1\overline{1}\rangle$ in Eq.~(\ref{cnotothermap}), while there is no way to realize such an effect in Eq.~(\ref{czlikemap}). This last effect basically means that the mechanism for realizing Eq.~(\ref{cnotothermap}) can be used to realize a high-fidelity multi-qubit gate: one can use TSD to excite the input state $|1\overline{1}\rangle$ to $|\mathscr{S}_+\rangle$ which can block the Rydberg pumping in a nearby atom~(of course care shall be taken for the design of such a gate); notice that the three-qubit gate in Ref.~\cite{Levine2019} requires exciting the two edge atoms to block the Rydberg pumping in the middle atom which results in error due to the residual blockade between the two edge atoms. For the method in Eq.~(\ref{czlikemap}), detuned Rabi cycles are used and it can not excite {\it one} Rydberg excitation in the blocked excitation~[for the input state $|11\rangle$ in Eq.~(\ref{czlikemap})] , and thus it is not possible to extend the method in Eq.~(\ref{czlikemap}) to a high-fidelity multi-qubit gate. This means that the underlying physics in Eq.~(\ref{cnotothermap}) is different from that of Eq.~(\ref{czlikemap}).

Note that the basis transform used to write the TSD-based {\footnotesize CNOT} in Eq.~(\ref{czlikemap}) is used only for clarifying the physics, but not for practical use since in a large-scale quantum computer the way to encode quantum information shall be based on a commonly used qubit basis in all registers~(otherwise the information loading and retrieving will face trouble). For example, to initialize the atomic arrays, the quantization axis is fixed by applying, e.g., an external magnetic field, and the qubit states $|0\rangle$ and $|1\rangle$ are chosen from two hyperfine levels with different energy. So, the states $|\overline{1}(\overline{0})\rangle=(|0\rangle\pm|1\rangle)/\sqrt{2}$ are no longer eigenstates because of the applied external fields, and new state detection schemes shall be designed. In a large-scale quantum computer, to play the role of control or target is not fixed for any qubit: a computation task is divided into a series of unitary operations, in different unitary operations, the same qubit sometimes serves as a control, and sometimes as a target. This means that if one uses a hybrid coding with $|0(1)\rangle$ for the control and $|\overline{1}(\overline{0})\rangle$ for the target, an exceedingly large amount of transforming operations between $|0(1)\rangle$ and $|\overline{1}(\overline{0})\rangle$ should be used in the quantum circuit. So, it is impractical to use $|0(1)\rangle$ for information coding in the control, and to use $|\overline{1}(\overline{0})\rangle$ for information coding in the target; for a proof-of-principle study either in theory or in experiment, it may be of interest, but not for a realistic large-scale quantum computer.

\section{AC Stark shifts}\label{appendixE}
There is a detailed study about the ac Stark shifts in a two-photon Rydberg excitation in Ref.~\cite{Maller2015}. Appendix B of~\cite{Maller2015} presents a detailed calculation of the ac Stark shifts accounting for the hyperfine splitting of the intermediate state, where one can find that compensation of the ac Stark shifts is feasible. Below, we ignore the hyperfine splitting of the intermediate state to give a brief introduction. More details can be found in Ref.~\cite{Maller2015}.

Here, we choose the qubit states by $|1\rangle=|6S_{1/2},F = 3,m_F = 3\rangle$ and $|0\rangle=|6S_{1/2},F = 4,m_F = 4\rangle$ of cesium. The state $|1\rangle$ is driven to $|p\rangle$~(in the $7P_{1/2}$ manifold) with laser fields that are left-hand circularly polarized~(note that this is a little different from Ref.~\cite{Maller2015}), which is further excited to $|r\rangle=|82S_{1/2},m_J,m_I\rangle$. The Hamiltonian for a two-photon Rydberg excitation is~(in this section we explicitly put $\hbar$ in Hamiltonians)
\begin{eqnarray}
  \hat{H}_{\text{2-pho;0}} &=&\hbar [\Omega_1 |p\rangle\langle 1|/2+ \Omega_2 |r\rangle\langle p|/2+\text{H.c.}]\nonumber\\ &&+\hbar \Delta |p\rangle\langle p|,\label{equation02}
\end{eqnarray}
where $\Delta$ is defined as the frequency of the laser field deducted by the frequency of the atomic transition. Note that we have not included the nonresonant shift in the equation above. When $\Delta$ is very large compared to the decay rate of $|p\rangle$, the intermediate state can be adiabatically eliminated, leading to 
\begin{eqnarray}
  \hat{H}_{\text{2-pho}} &=&\hbar\{ [\Omega_{\text{eff}} |r\rangle\langle 1|/2+ \text{H.c.}]+ \Delta_r |r\rangle\langle r|\nonumber\\ &&+ \Delta_{q1} |1\rangle\langle 1|\},  \label{equation03}
\end{eqnarray}
where $\Omega_{\text{eff}}=\Omega_1\Omega_2/(2\Delta)$, and according to Ref.~\cite{Maller2015} the effective detuning at the level $|r\rangle$ is
\begin{eqnarray}
 \Delta_r(\omega_1,\omega_2,\Delta,\mathcal{E}_1,\mathcal{E}_2)&=& \frac{\Omega_2^2}{4\Delta}-\frac{e^2}{4m_e\hbar}\left( \frac{\mathcal{E}_1^2}{\omega_1^2}+ \frac{\mathcal{E}_2^2}{\omega_2^2}  \right),\nonumber\\~\label{equation04}
\end{eqnarray}
where $e$ is the elementary charge, $m_e$ the mass of the electron, and $\omega_j$ and $\mathcal{E}_j$ are the frequency and electric field amplitude of the laser field, where $j=1$ and $2$ for the lower and upper transitions, respectively. Similarly, for the ground state, the effective detuning is
\begin{eqnarray}
 \Delta_{q1}(\omega_1,\omega_2,\Delta,\mathcal{E}_1,\mathcal{E}_2) &=& \frac{\Omega_1^2}{4\Delta}-\frac{1}{4\hbar}\left( \alpha_1 \mathcal{E}_1^2+ \alpha_2 \mathcal{E}_2^2  \right),\nonumber\\~\label{equationsec06eq13}
\end{eqnarray}
where $\alpha_1$ and $\alpha_2$ are the nonresonant polarizabilities from the lower and upper fields that can be calculated via the sum over transitions to other states except the intermediate state $|p\rangle$. Meanwhile, there is a shift on the ground state $|0\rangle$
\begin{eqnarray}
  \Delta_{q0}(\omega_1,\omega_2,\Delta,\mathcal{E}_1,\mathcal{E}_2) &=& \frac{\mathscr{C}^2\Omega_1^2}{4(\Delta+\omega_q)}-\frac{1}{4\hbar}( \alpha_1 \mathcal{E}_1^2\nonumber\\&&+ \alpha_2 \mathcal{E}_2^2  ),
  \label{equationsec06eq14}
\end{eqnarray}
where $\omega_q$ is the frequency separation between $|0\rangle$ and $|1\rangle$, and $\mathscr{C}$ is a factor determined by the selection rules. Because $\alpha_2/\alpha_1=-16.3$ and $\alpha_1$ is negative, the off-resonant shift will be negative, and thus it is necessary for the resonant shift to be positive. So, both $\Delta$ in Eq.~(\ref{equationsec06eq13}) and $\Delta+\omega_q$ in~(\ref{equationsec06eq14}) shall be positive, which further means that $\mathscr{C}^2$ shall be larger than 1.

For the case when $\Delta$ is much larger than the hyperfine splitting of the intermediate $7P_{1/2}$ state, we ignore its hyperfine splitting and then it can be written as $|7P_{1/2},m_J,m_I\rangle$. When right-hand polarized fields are used for coupling qubit states and the $7P_{1/2}$ states, the square of the ratio of the coupling between $|0\rangle$ and $|p\rangle$ and that between $|1\rangle$ and $|p\rangle$ is
\begin{eqnarray}
 \mathscr{C}^2 &=& \frac{\sum_{m_J,m_I} \left(\sum_{m_e}C_{m_J,1,m_e}^{1/2,1,1/2}C_{m_e,m_I,0}^{1/2,7/2,4}\right)^2}{\sum_{m_J,m_I} \left(\sum_{m_e}C_{m_J,1,m_e}^{1/2,1,1/2}C_{m_e,m_I,0}^{1/2,7/2,3}\right)^2},
\end{eqnarray}
which is 8. To shift $ \Delta_r(\omega_1,\omega_2,\Delta,\mathcal{E}_1,\mathcal{E}_2)$ to zero, one can simply adjust the frequency of the laser fields for addressing the upper transition. To compensate the Stark shifts at $|0\rangle$ and $|1\rangle$, Eqs.~(\ref{equationsec06eq13}) and~(\ref{equationsec06eq14}) indicates that $\mathscr{C}^2=1+\omega_q/\Delta$, which means that we shall choose $\Delta=\omega_q/( \mathscr{C}^2-1)\approx2\pi\times1.3$~GHz~(for rubidium it would be $2\pi\times0.97$~GHz) which is near to values used in experiments~\cite{Isenhower2010,Maller2015}. With $\Delta=\omega_q/( \mathscr{C}^2-1)$ satisfied, an appropriate set of $(\Delta,\mathcal{E}_1,\mathcal{E}_2)$ satisfying Eq.~(\ref{equationsec06eq13}) would satisfy Eq.~(\ref{equationsec06eq14}), too. With the given data for the radial coupling between ground and $|p\rangle$ states and the values of $\alpha_{1,2}$~(page 6 of~\cite{Maller2015}), Eq.~(\ref{equationsec06eq13}) reduces to $\overline{\Delta}[16.3\left(\frac{\mathcal{E}_2}{\mathcal{E}_1}\right)^2-1 ]=2.98$, where $\overline{\Delta}=\Delta\times10^{-9}s$, which can be used to set up the ratio $|\mathcal{E}_2/\mathcal{E}_1|$.

We have shown the method to compensate the Stark shifts if pumping applies only to the transition $|1\rangle\leftrightarrow|r\rangle$. For the target qubit, there will be another transition required, $|0\rangle\leftrightarrow|r\rangle$. Then, four fields lead to Stark shifts in the two qubit states~(two equations), and there is a condition that the Rabi frequencies for $|0(1)\rangle\leftrightarrow|r\rangle$ shall be equal~(a third equation). Moreover, there will be two detunings at the intermediate states for these two transitions. Altogether there are three equations for six variables and hence the problem is solvable. The Stark shift at $|r\rangle$ can be compensated by adjusting the central frequencies of the laser fields. So, Eq.~(\ref{SM03}), or Eq.~(3) in the main text, can be realized.  

For one-photon excitation of $p$-orbital Rydberg states, the off-resonant ac Stark shift on $|0\rangle$ or $|1\rangle$ from the resonant field is positive, but one can use another field of larger wavelength to induce a negative shift. Then, a similar set of equations like  Eqs.~(\ref{equationsec06eq13}) and~(\ref{equationsec06eq14}) can be established to compensate the Stark shifts. For the shift on the Rydberg state, one can shift the wavelength of the laser field to recover the resonance condition for $|r\rangle$.

\section{{\footnotesize CNOT} in other systems}\label{appendixF}
It is useful to compare the speed of {\footnotesize CNOT} in different physical systems by using the ratio $\Xi=T_{\text{coh}}/T_{\text{g}}$, where $T_{\text{coh}}$ is the coherence time~(pure spin dephasing $T_2$ or the inhomogeneously broadened $T_2^\ast$), and $T_{\text{g}}$ is the duration of {\footnotesize CNOT}. First, for neutral atom qubit systems, in Ref.~\cite{Wang2016} a coherence time of 7 seconds has been measured (see the left column of page 2 of~\cite{Wang2016}) in a large neutral atom array. Ref.~\cite{Wang2016} realized such a coherence time so as to have high fidelity in the single-qubit gates that had gate durations~80~$\mu$s; in principle, much longer coherence times can be achieved. To have a conserved estimation, we assume that a coherence time of 7s for neutral atoms. With the protocol in this manuscript, the {\footnotesize CNOT} duration would be about 0.3~$\mu$s with Rydberg Rabi frequencies like those realized in Refs. [43, 44]. Then, the figure-of-merit would be $\Xi=2.3\times10^7$.

Second, for trapped ions, several most recent and most advanced results can be inferred from literatures. However, one should bare in mind that some fast entangling gates are not {\footnotesize CNOT} gates. Any two-qubit entangling gate can be repeated several times to form a {\footnotesize CNOT} when assisted by single-qubit gates, as demonstrated in~\cite{Bremner2002}, but the {\footnotesize CNOT} is the gate directly helpful to quantum computation in the circuit model~\cite{Shor1997,Bremner2002,Shende2004,Peruzzo2014,Debnath2016}. So, we focus on data for trapped ions with {\footnotesize CNOT} gates demonstrated. (i) In Ref.~\cite{Ballance2015}, Bell states of trapped ions were created with gate times 27.4~$\mu$s. The coherence times were not mentioned in Ref.~\cite{Ballance2015}. However, near the end of Ref.~\cite{Ballance2015} it stated coherence times of 60s which indicates that the coherence times of their qubits were around 60s. So the figure-of-merit would be $60s/(27.4~\mu$s) which is about $\Xi=2.2\times10^6$. (ii) In Ref. ~\cite{Ballance2016}, Bell states of trapped ions were prepared by pulses with durations from 50$\mu$s to 100$\mu$s; it also showed very short gate times of about 3.8$\mu$s. In the Supplemental Material of Ref. ~\cite{Ballance2016}, it showed that the coherence time is about 6s. So the figure-of-merit is 6s/(3.8$\mu$s) which is about $\Xi=1.6\times10^6$. (iii) In Ref.~\cite{Gaebler2016}, Bell states of trapped ions were prepared by pulses of duration about 30$\mu$s~(Fig. 6 of Ref. \cite{Gaebler2016} showed gates with durations even longer), and the coherence time is about 1.5s. So the figure-of-merit would be 1.5s/(30$\mu$s) which is about $\Xi=5\times10^4$. 

Third, for superconducting qubits, very fast {\footnotesize CNOT} gates were reported in Ref.~\cite{Barends2014}. The energy relaxation time~$T_1$ is usually shorter than the phase coherence time~$T_2^\ast$ in superconducting circuits. In~\cite{Barends2014} the measured value of $T_1$ was $20-40~\mu$s and values up to $57~\mu$s were recorded. The Supplementary Information of Ref.~\cite{Barends2014} showed that the $C_Z$ gate durations can be as short as $38$~ns, and the single-qubit gate times can be as small as $10$~ns. So the gate duration of a {\footnotesize CNOT} can be $T_{\text{g}}=58$~ns, leading to a figure-of-merit $\Xi=10^3$. Suppose the fast gate in Ref.~\cite{Barends2014} was realized in the superconducting system with a longer coherence time $T_1=70~\mu$s~\cite{Rigetti2012}, the figure-of-merit $\Xi=1.2\times10^3$ is still much smaller than those in neutral atoms.

Concerning the ratio between the coherence time and the {\footnotesize CNOT} duration, the above comparison shows that TSD makes neutral atoms advantageous compared to trapped ions and superconducting systems.


%


\end{document}